\documentclass[12pt,a4]{article}

\usepackage{amsthm,amsmath,latexsym,amssymb,amsfonts,amscd}
\usepackage{graphics,lscape,fancyhdr,array,stmaryrd,euscript}
\usepackage{empheq}
\usepackage{dsfont}
\usepackage{verbatim}
\usepackage{color,tikz,tikz-cd}
\usetikzlibrary[snakes]
\usepackage{relsize,slashed}
\numberwithin{equation}{section}
\usepackage{hyperref}

\newcommand{\bep}{\begin{picture}}
\newcommand{\eep}{\end{picture}}

\newcommand{\smallpic}[1]{{\unitlength=0.2mm#1}}

\newcounter{YoungHeight}\newcounter{YoungWidth}

\newcounter{Mul1}\newcounter{Mul2}\newcounter{Mul3}\newcounter{Mul4}
\newcounter{A0}\newcounter{A1}\newcounter{A2}
\newcounter{B3}
\newcounter{C3}\newcounter{C4}
\newcounter{D1}\newcounter{D2}\newcounter{D3}
\newcounter{T0}\newcounter{T1}

\newlength{\txtHShift}

\newlength{\txtWidth}

\newcommand{\HalfLength}[2]{\setcounter{Mul1}{#1}\setcounter{Mul2}{#1}\addtocounter{Mul1}{\value{Mul2}}\addtocounter{Mul1}{\value{Mul2}}%
\addtocounter{Mul1}{\value{Mul2}}\addtocounter{Mul1}{\value{Mul2}}\setcounter{#2}{\value{Mul1}}}

\newcommand{\Add}[3]{\setcounter{#1}{#2}\addtocounter{#1}{#3}}

\newcommand{\Length}[1]{#10}
\newcommand{\YoungScale}{}

\newcommand{\shiftedText}[2]{{\hspace{#1}#2}}
\newcommand{\calcHShift}[1]{\settowidth{\txtWidth}{#1}\setlength{\txtHShift}{-0.5\txtWidth}}

\newcommand{\TextTop}[3]{{\calcHShift{#1}\HalfLength{#2}{T0}\Add{T1}{\Length{#3}}{-9}\put(\value{T0},\value{T1}){\shiftedText{\txtHShift}{#1}}}}

\newcommand{\BlockA}[2]{{\YoungScale\bep(\Length{#1},\Length{#2}){\Add{A1}{#1}{1}\Add{A2}{#2}{1}}%
\multiput(0,0)(10,0){\value{A1}}{\line(0,1){\Length{#2}}}\multiput(0,0)(0,10){\value{A2}}{\line(1,0){\Length{#1}}}%
\setcounter{YoungHeight}{\Length{#2}}\setcounter{YoungWidth}{\Length{#1}}\eep}}

\newcommand{\RectT}[3]{\bep(\Length{#1},\Length{#2})\put(0,0){\line(1,0){\Length{#1}}}\put(0,0){\line(0,1){\Length{#2}}}%
\put(\Length{#1},\Length{#2}){\line(-1,0){\Length{#1}}}\put(\Length{#1},\Length{#2}){\line(0,-1){\Length{#2}}}#3{#1}{#2}\eep}

\newcommand{\RectBRow}[4]{{\bep(\Length{#1},20)\put(0,0){\RectT{#2}{1}{\TextTop{#4}}}%
\put(0,10){\RectT{#1}{1}{\TextTop{#3}}}\eep}}

\newcommand{\YoungA}{\BlockA{1}{1}}

\newcommand{\YoungAA}{\BlockA{1}{2}}

\newcommand{\BlockApar}[2]{\parbox{\Length{#1}pt}{\YoungScale\bep(\Length{#1},\Length{#2}){\Add{A1}{#1}{1}\Add{A2}{#2}{1}}%
\multiput(0,0)(10,0){\value{A1}}{\line(0,1){\Length{#2}}}\multiput(0,0)(0,10){\value{A2}}{\line(1,0){\Length{#1}}}%
\setcounter{YoungHeight}{\Length{#2}}\setcounter{YoungWidth}{\Length{#1}}\eep}}

\newcommand{\BlockBpar}[4]{\parbox{\Length{#1}pt}{\YoungScale\Add{B3}{\Length{#2}}{\Length{#4}}%
\bep(\Length{#1},\value{B3})\put(0,\Length{#4}){\BlockA{#1}{#2}}%
\put(0,0){\BlockA{#3}{#4}}\setcounter{YoungHeight}{\value{B3}}\setcounter{YoungWidth}{\Length{#1}}\eep}}

\newcommand{\YoungpA}{\BlockApar{1}{1}}
\newcommand{\YoungpB}{\BlockApar{2}{1}}

\newcommand{\YoungpD}{\BlockApar{4}{1}}

\newcommand{\YoungpAA}{\BlockApar{1}{2}}

\newcommand{\YoungpBB}{\BlockApar{2}{2}}
\newcommand{\YoungpCA}{\BlockBpar{3}{1}{1}{1}}

\newcommand{\BlockAm}[2]{\mbox{\smallpic{\YoungScale\bep(\Length{#1},\Length{#2}){\Add{A1}{#1}{1}\Add{A2}{#2}{1}}%
\multiput(0,0)(10,0){\value{A1}}{\line(0,1){\Length{#2}}}\multiput(0,0)(0,10){\value{A2}}{\line(1,0){\Length{#1}}}%
\setcounter{YoungHeight}{\Length{#2}}\setcounter{YoungWidth}{\Length{#1}}\eep}}}

\newcommand{\YoungmAA}{\raisebox{-2pt}{\BlockAm{1}{2}}}

\usepackage{setspace}

 \usepackage[numbers,sort&compress]{natbib}
 \usepackage[nottoc]{tocbibind}
\newcommand{\pl}{\partial}
\newcommand{\besubeqs}{\begin{subequations}}
\newcommand{\esubeqs}{\end{subequations}}

\newcommand{\mm}{{\ensuremath{{\mu}}}}
\newcommand{\nn}{{\ensuremath{{\nu}}}}

\newcommand{\ga}{\alpha}
\newcommand{\gb}{\beta}

\newcommand{\fud}[2]{{}^{#1}{}_{#2}\,}
\newcommand{\fdu}[2]{{}_{#1}{}^{#2}\,}

\newcommand{\Tr}{{\mathrm{Tr}\,}}

\newcommand{\PPP}{{\boldsymbol{P}}}
\newcommand{\LLL}{{\boldsymbol{L}}}
\newcommand{\KKK}{{\boldsymbol{K}}}
\newcommand{\DDD}{{\boldsymbol{D}}}
\newcommand{\TTT}{{\boldsymbol{T}}}

\newcommand{\ttt}{{\boldsymbol{t}}}
\newcommand{\SSS}{{\boldsymbol{S}}}
\newcommand{\RRR}{{\boldsymbol{R}}}

\newcommand{\hs}{\mathfrak{hs}}

\renewcommand{\d}{\partial}

\begin{document}
\pagenumbering{gobble}
\hfill
\vspace{-1.5cm}
\vskip 0.05\textheight
\begin{center}
{\Large\bfseries 
New Conformal Higher Spin Gravities in $\boldsymbol{3d}$}

\vspace{0.4cm}

\vskip 0.03\textheight

Maxim \textsc{Grigoriev}${}^{a,b}$, Iva \textsc{Lovrekovic}${}^c$ and Evgeny \textsc{Skvortsov}${}^{a,d}$

\vskip 0.03\textheight

\vspace{5pt}
{\em
$^a$ Lebedev Institute of Physics, \\
Leninsky ave. 53, 119991 Moscow, Russia\\

\vspace{5pt}

$^b$
Institute for Theoretical and Mathematical Physics,\\
Lomonosov Moscow State University, 119991 Moscow, Russia
\\
\vspace{5pt}
$^c$ Theoretical physics group, Blackett Laboratory,\\ Imperial College London, SW7 2AZ, U.K.\\
\vspace{5pt}
$^d$  Albert Einstein Institute, \\
Am M\"{u}hlenberg 1, D-14476, Potsdam-Golm, Germany
}
\end{center}

\vskip 0.02\textheight

\begin{abstract}
We propose a new class of conformal higher spin gravities in three dimensions, which extends the one by Pope and Townsend. The main new feature is that there are infinitely many examples of the new theories with a finite number of higher spin fields, much as in the massless case. The action has the Chern-Simons form for a higher spin extension of the conformal algebra. In general, the new theories contain Fradkin-Tseytlin fields with higher derivatives in the gauge transformations, which is reminiscent of partially-massless fields. A relation of the old and new theories to the parity anomaly is pointed out.
\end{abstract}

\newpage
\section{Introduction and Main Results}
\pagenumbering{arabic}
\setcounter{page}{2}
As is well-known, in three dimensions the Einstein-Hilbert action is equivalent \cite{Achucarro:1987vz,Witten:1988hc}, up to the known subtleties, to the Chern-Simons action for $so(2,2)\sim sl_2\oplus sl_2$. Its higher spin extension is obtained \cite{Blencowe:1988gj,Bergshoeff:1989ns,Campoleoni:2010zq,Henneaux:2010xg} by replacing $sl_2$ with any algebra such that its decomposition with respect to a given $sl_2$ embedding leads to representations bigger than the adjoint one. Even though the graviton, $s=2$, and massless higher spin fields, $s>2$, do not have propagating degrees of freedom in three dimensions this class of models has been very useful in understanding many aspects of higher spin theories and AdS/CFT, see e.g. \cite{Gaberdiel:2012uj} for a review.   

Similarly, conformal gravity in three dimensions can be understood as the Chern-Simons theory for the conformal algebra $so(3,2)$ \cite{vanNieuwenhuizen:1985cx,Horne:1988jf}. Conformal higher spin algebras in three dimensions are exactly the anti-de Sitter higher spin algebras in four dimensions. The Chern-Simons action for any of these algebras leads to a consistent theory of conformal higher spin fields. The first example of such a theory was studied by Pope and Townsend long ago \cite{Pope:1989vj} and by Fradkin and Linetsky \cite{Fradkin:1989xt}. They lead to a nonlinear theory of conformal higher spin fields, Fradkin-Tseytlin fields \cite{Fradkin:1985am}, that can be realized as symmetric traceless tensors $\phi_{a_1...a_s}$ with the linearized gauge symmetries of the form
\begin{align}\label{usualFT}
    \delta \phi_{a_1...a_s} &= \pl_{a_1}\xi_{a_2...a_s}+\text{permutations}-\text{traces}
\end{align}
Our proposal is to take advantage of the large class of higher spin algebras available from the study of symmetries of (higher derivative) free CFT's of type $\square^k\Phi(x)=0$ \cite{Eastwood2008,Michel,GoverSilhan,Bekaert:2013zya} and partially-massless fields in $AdS_{4}$  \cite{Dolan:2001ih,Bekaert:2013zya,Alkalaev:2014nsa,Basile:2014wua} and to construct some new algebras that are finite-dimensional \cite{Boulanger:2011se,Joung:2015jza}. This leads to several observations: (i) there exists a new class of nonlinear conformal higher spin theories in three dimensions; (ii) among them there are theories with finitely many fields in contrast to the conformal higher spin theories studied in \cite{Pope:1989vj,Fradkin:1989xt}; (iii) there is a large class of new higher spin algebras available. 

In addition to the usual Fradkin-Tseytlin fields \eqref{usualFT} the spectrum of the new conformal higher spin theories also involves conformal cousins \cite{Deser:1983tm,Erdmenger:1997wy,Vasiliev:2009ck, Bekaert:2013zya, Beccaria:2015vaa, Kuzenko:2019ill} of partially-massless fields \cite{Deser:1983tm,Higuchi:1986py,Deser:2001us,Zinoviev:2001dt, Skvortsov:2006at} that have higher derivatives in the gauge transformations:
\begin{align}\label{FTdepth}
    \delta \phi_{a_1...a_s} &= \pl_{a_1}...\pl_{a_t}\xi_{a_{t+1}...a_s}+\text{permutations}-\text{traces}
\end{align}
These fields can also be understood as boundary values of the genuine partially-massless fields in $AdS_4$ \cite{Bekaert:2013zya}. The complete theories for conformal higher spin fields are given by the Chern-Simons action for a variety of higher spin extensions of the conformal algebra $so(3,2)$.

What is interesting as compared to the case studied by Pope and Townsend is that there is a family of finite-dimensional higher spin algebras, which is similar to the purely massless case in $3d$. The theories with finitely many higher spin fields should be more tractable as the massless case has already shown. In the latter case the key role is played by $sl_2$ and its embedding into a higher spin algebra. Likewise, in the conformal case the key role is played by $so(3,2)$ and its embedding. In brief, one can pick any representation $V$ of $so(3,2)$, whether finite-dimensional or not. There is an associated higher spin algebra $\hs(V)=\mathrm{End}(V)$ that decomposes into $so(3,2)$-modules as $V\otimes V^*$. 

The $so(3,2)$ decomposition of $\hs(V)$ determines the spectrum of conformal fields, which is very similar to the partially-massless case \cite{Skvortsov:2006at}. Indeed, given a higher spin algebra $\hs$, the spectrum of the theory features a depth-$t$ spin-$s$ Fradkin-Tseytlin field \eqref{FTdepth} if the $so(3,2)$ decomposition of $\hs$ contains an irreducible module corresponding to the two-row Young diagram with rows of lengths $s-1$, $s-t$, where $t=1,...,s$:
\begin{align}
    \delta \phi_{a_1...a_s} &= \pl_{a_1}...\pl_{a_t}\xi_{a_{t+1}...a_s}+... && \Longleftrightarrow && \parbox{90pt}{\begin{picture}(90,20)\put(0,0){\RectBRow{9}{7}{$s-1$}{$s-t$}}\put(0,0){\YoungAA}\put(60,0){\YoungA}\put(80,10){\YoungA}\end{picture}}
\end{align}
This dictionary is established by studying the free approximation and the Chern-Simons action provides a simple and apparently unique non-linear completion.

While the Chern-Simons Lagrangian is simple and manifestly gauge-invariant up to a total derivative, an attempt to solve for the auxiliary fields and write down a closed form expression in terms of the conformal higher spin fields $\phi_{a_1...a_s}$ faces great technical difficulties \cite{Nilsson:2013tva,Nilsson:2015pua,Linander:2016brv}. The same is true for massless fields in three dimensions --- the Chern-Simons formulation is by far more economical and simple at the moment \cite{Fredenhagen:2014oua,Campoleoni:2012hp}. Nevertheless, it is worth stressing that conformal fields $\phi_{a_1...a_s}$ or its massless analogs, Fronsdal fields, can easily be shown to be the only dynamical variables present in the Chern-Simons connection, with the rest of the components being (generalized) auxiliary fields. The equations of motion have the form of vanishing (higher spin generalization of) Cotton tensor:
\begin{align}
    C_{a_1...a_s}&=\epsilon^{...}\pl^{2s-2t+1} \phi_s +\mathcal{O}(\phi^2)\,,
\end{align}
where the expression is sketchy and shows the order of derivatives, the $\epsilon^{abc}$-tensor and the presence of higher order corrections that involve fields of different spins (specified by the higher spin algebra). For the usual $t=1$ case the linearized Cotton tensor was discussed in \cite{Pope:1989vj,Henneaux:2015cda,Kuzenko:2016qwo,Kuzenko:2016qdw,Basile:2017mqc,Buchbinder:2018dou,Kuzenko:2019ill,Buchbinder:2019yhl} for a number of cases and certain nonlinear terms were derived in \cite{Nilsson:2013tva,Nilsson:2015pua,Linander:2016brv}. For $t>1$ the linearized Cotton tensors were derived in \cite{Kuzenko:2019ill}.

Conformal higher spin fields in $d>3$ are non-unitary, but this is not the case in three dimensions due to the topological nature of conformal fields. Therefore, conformal higher spin fields in three dimensions are as good as massless fields, while the unitarity is clearly an issue in $d>3$. Historically, the $t>1$ Fradkin-Tseytlin fields were studied in few papers \cite{Deser:1983tm, Erdmenger:1997wy, Vasiliev:2009ck, Bekaert:2013zya, Beccaria:2015vaa, Barnich:2015tma, Metsaev:2016oic, Kuzenko:2019ill}. The $d=3$ case is very much different from the $d>3$ ones since the conformally-invariant differential equation is given by the Cotton tensor \cite{Kuzenko:2019ill}. 

In the paper we discuss the conformal higher spin theories as the Chern-Simons theory for appropriate higher spin algebras. Without going into detail let us mention another possibility to construct such theories as effective actions. It is known that the conformal higher spin gravity in four dimensions \cite{Segal:2002gd,Tseytlin:2002gz,Bekaert:2010ky} (any even dimension) is closely related to the conformal anomaly  --- it can be defined as the local part of the induced action, the coefficient of the $\log$-divergent term \cite{Tseytlin:2002gz}. In three dimensions there also exists a way to obtain the conformal higher spin gravities as parity anomalies. Indeed, it is well-known that coupling fermions to gauge fields induces the parity anomaly \cite{Niemi:1983rq,Redlich:1983dv,Redlich:1983kn}
\begin{align}
    \mathrm{Im} \log \int D\psi D\bar{\psi}\,  e^{i \int d^3x\, [i\bar\psi \slashed\partial\psi +\bar\psi \gamma_\mu \psi A^\mu]} &=   S_{CS}[A]
\end{align}
and the anomaly is given by the Chern-Simons action (we do not display its level). This anomaly results from coupling of the spin-one current $\bar{\psi} \gamma_\mu \psi$ to the background field $A^\mu$. A similar result can be obtained for the gravitational coupling \cite{AlvarezGaume:1984nf} and for four-manifolds with boundaries \cite{Kurkov:2018pjw}. The free fermion theory in three dimensions features conserved tensors of any rank $s=1,2,3,4,....$, which are schematically of the form $J_s=\bar \psi \gamma \pl^{s-1}\psi$. The tensor currents $J_s$ are conserved and traceless on-shell. Therefore, the corresponding background fields are exactly the Fradkin-Tseytlin fields $\phi^{a_1...a_s}$ discussed above, which extends $A^\mu$ and $g^{\mu\nu}$ to any spin. Since the currents associated with $J_s$ can also be represented as on-shell closed two-forms, the natural object they couple to is a connection $\omega$ of the higher spin algebra \cite{Didenko:2012vh}. It should not be hard to show that the conformal higher spin gravity of Pope and Townsend can be obtained as the parity anomaly
\begin{align}
    \mathrm{Im} \log \int D\psi D\bar{\psi}\,  e^{i \int d^3x\,[i\bar\psi \slashed\partial\psi+\sum_s J_{a_1...a_s}\phi^{a_1...a_s}]} &=  S_{CS}[\omega]
\end{align}
see \cite{Bonora:2016ida} for the first steps in this direction and related discussion. Note that the relation between connection $\omega$ and Fradkin-Tseytlin fields $\phi^{a_1...a_s}$ is quite complicated and is not known explicitly, in general.

At least some of the new theories that we discuss in the paper should correspond to the parity anomaly in the higher derivative theories with conformally-invariant kinetic terms of the form $\bar\psi\square^k(\slashed\partial)\psi$. We also note that the parity anomaly can be used to define conformal higher spin theories in all odd dimensions $d$, the action being the $d$-dimensional Chern-Simons term. It would also be interesting to see what is the higher spin generalization of the Atiyah-Patodi-Singer theorem \cite{Atiyah:1975jf}.

We expect that the new theories provide a useful higher spin playground due to both the large family of new examples and existence of theories with finitely many fields on top of the conformal graviton. The supersymmetric extensions of the new higher spin theories should also exist and be based on higher spin extensions of $osp(N|4)$, \cite{Fradkin:1989xt}.
The finite dimensional examples may help to establish a precise relation between the Chern-Simons formulation and nonlinear completion of the Fradkin-Tseytlin fields. It would also be interesting to explore the relation between the parity anomaly and conformal higher spin theories. Given that conformal gravity admits both anti-de Sitter and flat backgrounds, another useful application is to investigate various aspects of gauge/gravity duality, extending \cite{Afshar:2011qw, Bagchi:2012yk,Afshar:2013bla} to the new conformal theories.

The outline of the paper is simple. For the reader's convenience we review the Chern-Simons formulation of the conformal gravity in section \ref{sec:review}. In section \ref{sec:newchs} we construct new conformal higher spin gravities, give a list of the relevant higher spin algebras and work out the dynamical content in terms of Fradkin-Tseytlin fields. In section \ref{sec:background} we discuss conformal higher spin fields as background fields for (topological) matter. 

\section{Conformal Gravity}
\label{sec:review}
We briefly discuss the conformal gravity in $3d$: how the simple Chern-Simons formulation is related to the one where the dynamical field is conformal graviton $g_{\mu\nu}$ \cite{Horne:1988jf}. The very first formulation of conformal gravity in three dimensions was found in the form of a non-standard Chern-Simons action \cite{Deser:1981wh,Deser:1982vy}:
\begin{align}\label{csintermsofe}
    S[e]&=\int \Tr\left[\varpi \wedge d \varpi +\frac23 \varpi\wedge \varpi\wedge \varpi\right]\,,
\end{align}
where\footnote{Indices $a,b,...=0,1,2$ are the indices of the $3d$ Lorentz algebra $so(2,1)$. Indices of differential forms are denoted $\mm,\nn,...=0,1,2$. } $\varpi\equiv \frac12 \varpi^{a,b} \LLL_{ab}$. Here $\varpi^{a,b}=\varpi^{a,b}_\mm\, dx^\mm$ is a spin-connection and $\LLL_{ab}$ are the generators of the Lorentz algebra $so(2,1)$ with the canonical trace in the adjoint representation. However, the spin-connection $\varpi$ is not an independent variable, rather it is expressed in terms of dreibein $e^a\equiv e^a_\mm \, dx^\mm$ via the torsion constraint
\begin{align}
    \nabla e^a&= de^a+\varpi\fud{a,}{b}\wedge e^b=0\,.
\end{align}
Also, the conformal invariance of the action is not manifest. This formulation is equivalent to the Chern-Simons theory of the conformal algebra $so(3,2)$ \cite{Horne:1988jf}. To demonstrate this let us fix the commutation relations of $so(3,2)$ to be 
\besubeqs\label{LPDK}
\begin{align}
[\DDD,\PPP^a]&=-\PPP^a\,, & [\LLL^{ab},\PPP^c]&=\PPP^a\eta^{bc}-\PPP^b\eta^{ac}\,,\\
[\DDD,\KKK^a]&=+\KKK^a\,,  & [\LLL^{ab},\KKK^c]&=\KKK^a\eta^{bc}-\KKK^b\eta^{ac}\,,\\
[\PPP^a,\KKK^b]&=-\LLL^{ab}+\eta^{ab}\DDD\,, &[\LLL^{ab},\LLL^{cd}]&=\LLL^{ad}\eta^{bc}+\text{three more}\,,
\end{align}
\esubeqs
where $\LLL^{ab}$, $\PPP^a$, $\KKK^a$ and $\DDD$ are Lorentz, translation, conformal boosts and dilation generators, respectively. Now, we take a connection of $so(3,2)$
\begin{align}
    \omega&= \frac12 \varpi^{a,b} \LLL_{ab}+e^a\PPP_a +f^a \KKK_a+ b\DDD
\end{align}
and write down the Chern-Simons action for $\omega$ (we drop the level)
\begin{align}\label{csforso32}
    S[\omega]&=\int \Tr\left[\omega \wedge d \omega +\frac23 \omega\wedge \omega\wedge \omega\right]\,.
\end{align}
The equations set the curvature $F=d\omega+\frac12 [\omega,\omega]$ to zero. The component form reads
\besubeqs
\begin{align}
    F_P^{a}&=\nabla e^a- b\wedge e^a\,,\\
    F_D&= \nabla b+ e_m \wedge f^m\,,\\
    F_L^{a,b}&= R^{a,b}-e^a\wedge f^b+e^b\wedge f^a\,,\\
    F_K^a&=\nabla f^a+b\wedge f^a\,,
\end{align}
\esubeqs
where $\nabla =d+\varpi$ is the Lorentz covariant derivative; the Riemann two-form is $R^{a,b}=d\omega^{a,b}+\omega\fud{a,}{c}\wedge \omega^{c,b}$. In order to prove the equivalence we need to impose certain gauge conditions and solve some of the equations. The standard gauge transformations with gauge parameter 
\begin{align}
    \Xi&= \frac12 \eta^{a,b} \LLL_{ab}+\xi^a\PPP_a +\zeta^a \KKK_a+ \rho\DDD
\end{align}
leads to
\besubeqs
\begin{align}
    \delta e^a&=\nabla \xi^a+\eta\fud{a,}{b} \wedge e^a  - \rho\wedge e^a-b\wedge \xi^a\,,\\\
    \delta b&= \nabla \rho +e_m \wedge \zeta^m+\xi_m \wedge f^m\,,\\
    \delta\omega^{a,b}&= \nabla \eta^{a,b}-e^a\wedge \zeta^b+e^b\wedge \zeta^a-\xi^a\wedge f^b+\xi^b\wedge f^a\,,\\
    \delta f^a&=\nabla  \zeta^a+ \eta\fud{a,}{b} \wedge f^a+b\wedge \zeta^a+\rho\wedge f^a\,.
\end{align}
\esubeqs
Firstly, assuming that the dreibein $e^a_\mm$ is non-degenerate, we can gauge away $b$, i.e. impose $b=0$. Now, the gauge transformations of $e^a$ acquire the required form, so that one finds the correct transformations for the conformal metric $g_{\mu\nu}=e^a_\mu e^b_\nu \eta_{ab}$. Equation $F_D=0$ implies that $f^a_\mu e_{a\nu}$ is symmetric. $F_P^a=0$ is the standard torsion constraint to be solved for $\varpi^{a,b}$ in terms of $e^a$. Next, $F_L^{a,b}=0$ implies that $f^a_\mu e_{a\nu}$ is the Schouten tensor. Finally, $F_K^a=0$ imposes $C_{\mu\nu}=0$, where $C$ is the Cotton tensor. The latter is the only dynamical equation. 

If we impose $b=0$ and solve for all the auxiliary fields, i.e. $f^a$,  $\varpi^{a,b}$, and plug the solution back into the Chern-Simons action \eqref{csforso32} we obtain \eqref{csintermsofe}, where $e^a$ is the only dynamical variable. In fact, the action can be rewritten in terms of conformal metric $g_{\mu\nu}$.

\section{New Conformal Higher Spin Theories}
\label{sec:newchs}
As it was already stated in the introduction, the main idea is to use the large stock of higher spin algebras for (partially)-massless fields, which is available in $AdS_4$ and to add some new algebras. The same algebras can be interpreted as $3d$ conformal higher spin algebras. The action is the same Chern-Simons action, but for one of the algebras listed below:
\begin{align}
    S[\omega]&=\int \Tr\left[\omega \wedge d \omega +\frac23 \omega\wedge \omega\wedge \omega\right]\,.
\end{align}
Such a simple extension leads to more general Fradkin-Tseytlin fields with higher derivative gauge transformations and also to theories with a finite number of fields.

\subsection{Higher Spin Algebras}
\label{sec:hsa}
What is needed is a higher spin extension of $so(3,2)$, i.e. algebras that have $so(3,2)$ as a subalgebra and also contain various other nontrivial representations of $so(3,2)$. One such class is the partially-massless higher spin algebras. Originally, they were found as symmetries of conformally-invariant higher derivative equations $\square^k \Phi(x)=0$ \cite{Eastwood2008,GoverSilhan,Bekaert:2013zya}. From this point of view the usual massless higher spin algebra is a particular $k=1$ case of the partially-massless family. There are several useful definitions available in the literature: (i) via the universal enveloping algebra of $so(3,2)$ \cite{Eastwood2008,GoverSilhan,Bekaert:2013zya}; (ii) oscillator realizations \cite{Vasiliev:2003ev,Alkalaev:2014qpa,Alkalaev:2014nsa,Joung:2014qya,Joung:2015jza}; (iii) explicit structure constants \cite{Joung:2014qya, Joung:2015jza}. It was also found in various contexts \cite{Boulanger:2011se,Manvelyan:2013oua,Joung:2015jza} that there are certain finite dimensional algebras that can be viewed as higher spin algebras, at least up to some point.\footnote{It is unlikely that any of such algebras can lead to a consistent higher spin theory in $d>3$. } Below we extend this class of finite dimensional algebras. 

\paragraph{Finite-dimensional algebras. } In order to construct algebras with a finite spectrum of Fradkin-Tseytlin fields one can take any nontrivial  finite-dimensional irreducible representation $V$ of $so(3,2)$, i.e. an irreducible tensor or a spin-tensor. The next step is to evaluate $U(so(d,2))$ in $V$, i.e. multiply the generators\footnote{Indices $A,B,...=0,...,4$ are the indices of $so(3,2)$ with the invariant metric denoted $\eta_{AB}$.} $\TTT_{AB}=-\TTT_{BA}$ of $so(3,2)$ in this representation and identify the algebra, we denote $\hs(V)$, that they generate. This algebra is associative by construction. Since $V$ is irreducible the algebra is just $\mathrm{End}(V)$, i.e. the algebra of all matrices of size $\dim V$. Constructed this way $\hs(V)=\mathrm{End}(V)$ comes equipped with a specific embedding of $so(3,2)$ generators via $\TTT_{AB}$. The same algebra can be understood as the quotient $U(so(3,2))/I$ by a two-sided ideal $I$ that is the annihilator of $V$, i.e. $I$ contains all polynomials in $\TTT_{AB}$ that vanish on $V$. 

We are interested in the decomposition of $\hs(V)$ into irreducible $so(3,2)$-modules --- this decomposition, as is shown below, determines the spectrum of Fradkin-Tseytlin fields, which is very easy to compute in practice. To summarize, there are infinite-many higher spin extensions of $so(3,2)$ that are labelled by various irreducible $so(3,2)$-modules:\footnote{Various peculiarities of higher derivative theories of type $\square^k\Phi(x)=0$ were noticed in \cite{Brust:2016gjy}. It would be interesting to see if the finite-dimensional higher spin algebras can be explained from the CFT point of view. }
\begin{align}
    \hs(V) =\mathrm{End}(V)=V\otimes V^*\,.
\end{align}
Spin-tensor representations $V$ are also allowed.\footnote{For example, a toy-model was studied in \cite{Pope:1989vj}, where $\hs$ was taken to be $so(4,2)$. This can be understood as $sl(V)$ for $V$ being the spinor representation of $so(3,2)$. }
Further extensions can be obtained by tensoring $\hs(V)$ with the matrix algebra $\mathrm{Mat}_N$, which turns the otherwise abelian spin-one field associated with $1$ of $\mathrm{End}(V)$ into the Yang-Mills field.\footnote{One can also project onto the symmetric and anti-symmetric parts of the tensor product $V\otimes V$ (as a matter of fact $V\sim V^*$ for finite-dimensional representations of $so(3,2)$), the only restriction being for it to contain the adjoint, i.e. $so(3,2)$ itself. In this way one can define higher spin analogs of $so(V)$ and $sp(V)$. } Note, that $\hs(V)=gl(V)=sl(V)\oplus u(1)$ as a Lie algebra and the spin-one field always decouples unless the Yang-Mills groups are added. Since the finite-dimensional higher spin algebras result from $\mathrm{End}(V)$ they are equipped with a canonical trace operation $\Tr$, which allows us to write down the Chern-Simons action.  

As the simplest example, let us take the vector representation, which is denoted by one-cell Young diagram $\YoungpA$. The corresponding algebra, $\hs(\YoungpA)$, has the following spectrum:\footnote{When treated as a higher spin algebra in $AdS_4$ this spectrum corresponds to a spin-two field, partially-massless depth-$3$ spin-three field and to a spin-one field.}
\begin{align}
    \hs(\YoungpA)&=\YoungpA\otimes \YoungpA= \YoungpAA \oplus \YoungpB\oplus \bullet \,.
\end{align}
The algebra is just the matrix algebra with $(d+2)^2$ generators $\ttt\fdu{A}{B}$ decomposed with respect to $so(d,2)$ (we keep the discussion $d$-dimensional here, otherwise, $d=3$). The $gl_{d+2}$ commutation relations are
\begin{align}
    [\ttt\fdu{A}{B}, \ttt\fdu{C}{D}]&=-\delta\fdu{A}{D}\ttt\fdu{C}{B} +\delta\fdu{C}{B} \ttt\fdu{A}{D}\,.
\end{align}
In the $so(d,2)$ base, the irreducible generators are $\TTT_{AB}=-\TTT_{BA}$, $\SSS_{AB}=\SSS_{BA}$ and $\RRR$, i.e.
\begin{align}
    \TTT_{AB}&=\ttt_{A|B}-\ttt_{B|A}\,, & \SSS_{AB}&=\ttt_{A|B}+\ttt_{B|A}-\frac{2}{d+2}\ttt\fdu{C}{C}\,, &&\RRR=\ttt\fdu{C}{C}\,.
\end{align}
The commutation relations read 
\besubeqs
\begin{align}
    [\TTT_{AB},\TTT_{CD}]&= \eta_{BC} \TTT_{AD}-\eta_{AC} \TTT_{BD}-\eta_{BD} \TTT_{AC}+\eta_{AD} \TTT_{BC}\,,\\
    [\TTT_{AB},\SSS_{CD}]&= \eta_{BC}\SSS_{AD}-\eta_{AC}\SSS_{BD}+\eta_{BD}\SSS_{AC}-\eta_{AD}\SSS_{BC}\,,\\
    [\SSS_{AB},\SSS_{CD}]&= \eta_{BC}\TTT_{AD}+\eta_{AC}\TTT_{BD}+\eta_{BD}\TTT_{AC}+\eta_{AD}\TTT_{BC}\,,
\end{align}
\esubeqs
while $\RRR$ commutes with everything since it is associated with $1$ in $gl(V)$.

Next to the simplest example is to take the rank-two symmetric representation, denoted by $\YoungpB$, which leads to algebra $\hs(\YoungpB)$ \cite{Joung:2015jza} with generators $\ttt\fud{AB}{CD}$ that are symmetric and traceless in the upper and lower pairs of indices. Its $so(d,2)$ decomposition reads
\begin{align}
    gl(\YoungpB)&=\YoungpB\otimes \YoungpB= \YoungpBB \oplus \YoungpCA\oplus \YoungpD\oplus \YoungpB \oplus \YoungpAA \oplus \bullet\,.
\end{align}

\paragraph{Infinite-dimensional algebras. } A large class of infinite-dimensional algebras is given by the symmetries of (higher-order) singletons, i.e. free CFT's of type $\square^q\Phi=0$ and $\square^{q-1}\slashed \pl\Psi=0$, which were studied in \cite{Eastwood2008,GoverSilhan,Bekaert:2013zya,Alkalaev:2014nsa, Grigoriev:2014kpa,Joung:2015jza,Joung:2014qya,Basile:2014wua}. These algebras are more difficult to work with, but the structure constants are explicitly available \cite{Joung:2015jza,Joung:2014qya}. We only present the $so(3,2)$-spectrum, which reads
\begin{align}
  \square^q\Phi=0&: && \hs\Big|_{so(3,2)}=\bigoplus_{s=0}^{\infty}\bigoplus_{k=0}^{k=q-1} \parbox{90pt}{\RectBRow{9}{7}{$s-1$}{$s-2k-1$}}\,.
\end{align}
The invariant trace $\Tr$ on such algebras is known \cite{Joung:2015jza,Joung:2014qya}, which allows one to write down the Chern-Simons action.\footnote{Note that the (classical) Poisson limit of a higher spin algebra was taken in \cite{Pope:1989vj}. This allows one to write down equations, but not the action. Also, most of the interaction terms that are prescribed by the higher spin algebra disappear in the classical limit. The $3d$ conformal higher spin theory based on higher spin (super)-algebras was studied in \cite{Fradkin:1989xt}.} Note, however, that there are certain subtleties of the higher derivative theories in lower dimensions \cite{Brust:2016gjy}.\footnote{One interesting feature of $3d$ is that the higher spin algebras of $q=1$ scalar and fermion are isomorphic. It is tempting to conjecture that the same is true for $q>1$. For the $q=1$ case this fact is one of the hints that the three-dimensional bosonization duality takes place. } 

\subsection{Free Fradkin-Tseytlin Fields}
\label{sec:freefields}
Our aim is to identify the field content of the new conformal higher spin gravities. We take one of the higher spin algebras listed above. Let us denote it $\hs$. In order to read off the field content it is enough to linearize the theory over the Minkowski vacuum. In Cartesian coordinates this vacuum can be chosen as $\omega_0=h^a \PPP_a$, where $h^a\equiv h^a_\mm\, dx^\mm$ and the background dreibein $h^a_\mm$ is the unit matrix, $h^a_\mm=\delta^a_\mm$. We recall that $\PPP_a\in so(3,2)\subset \hs$ by construction. The free equations and the linearized gauge symmetries read
\begin{align}\label{freeeqs}
    d \omega+ \omega_0\wedge\omega+\omega\wedge \omega_0&=0\,, & \delta \omega&=d\xi+[\omega_0, \xi]\,.
\end{align}
Since the background $\omega_0$ occupies only the $so(3,2)$ subalgebra of $\hs$, the free equations decompose into a set of independent equations for each of the $so(3,2)$ irreducible modules that appear in the decomposition of $\hs$. Assuming that the generators of $\hs$ are $\ttt_\Lambda$, so that $\omega\equiv \omega^\Lambda\, \ttt_\Lambda$ and $\xi\equiv \xi^\Lambda\, \ttt_\Lambda$, equations \eqref{freeeqs} read
\begin{align}
    d\omega^\Lambda\, \ttt_\Lambda +h^a \wedge \omega^\Lambda\, [\PPP_a,\ttt_\Lambda]&=0\,, & \delta \omega^\Lambda\, \ttt_\Lambda&= d\xi^\Lambda\, \ttt_\Lambda +h^a \wedge \xi^\Lambda\, [\PPP_a,\ttt_\Lambda]\,.
\end{align}
It is clear that we need to create a dictionary between the Fradkin-Tseytlin fields and irreducible modules of $so(3,2)$ that can appear in $\hs$. In the most general case, a module in the decomposition of $\hs$ is spanned by an $so(3,2)$ tensor $\ttt_{C(k),D(m)}$ that has a symmetry of a two-row Young diagram\footnote{This means that $t$ is symmetric in $C_1...C_k$, it is also symmetric in $D_1...D_m$. It is traceless in any two indices. The Young symmetry condition is that the symmetrization of all $C_1...C_k$ with at least one index $D$ must vanish. To save space and time we sometimes abbreviate $C_1...C_k$ as $C(k)$. Also, symmetrization over all indices denoted by the same letter is implied, e.g. the Young condition is $V_{C(k),CD(m-1)}\equiv0$ and the sum over $k+1$ terms is implied here. We do not consider spin-tensors, but this can easily be done. }
\begin{align}
    & \begin{picture}(90,20)\put(0,0){\RectBRow{9}{7}{$\cdots k\cdots$}{$\cdots m \cdots$}}\put(0,0){\YoungAA}\put(60,0){\YoungA}\put(80,10){\YoungA}\end{picture}
\end{align}
The conformal algebra commutation relations are
\begin{align}\label{Tt}
\begin{aligned}
    [\TTT_{AB},\TTT_{CD}]&= \eta_{BC} \TTT_{AD}-\eta_{AC} \TTT_{BD}-\eta_{BD} \TTT_{AC}+\eta_{AD} \TTT_{BC}
\end{aligned}
\end{align}
and the action of $\TTT_{AB}$ on $\ttt_{C(k),D(m)}$ is the canonical one:
\begin{align}
\begin{aligned}
    [\TTT_{AB},\ttt_{C(k),D(m)}]&= \eta_{BC} \ttt_{AC(k-1),D(m)}-\eta_{AC} \ttt_{BC(k-1),D(m)}\\
    &+\eta_{BD} \ttt_{C(k),AD(m-1)}-\eta_{AD} \ttt_{C(k),BD(m-1)}
\end{aligned}\label{eq316}
\end{align}
The dictionary between \eqref{LPDK} and $\TTT_{AB}$ is $\PPP_a=\TTT_{a+}$, $\KKK_a=\TTT_{a-}$, $\DDD=-\TTT_{+-}$, $\LLL_{ab}=\TTT_{ab}$, where we chose the light-cone coordinates with $\eta_{+-}=\eta_{-+}=1$ and $A={a,+,-}$, etc. As a result, we can easily read off $[\PPP_a,\ttt_\Lambda]$ for $\ttt_\Lambda$ in any irreducible $so(3,2)$-module. Let us consider some simple examples.

\paragraph{Conformal Gravity.} First of all, it is instructive to redo the analysis for the conformal gravity case, which corresponds to the adjoint module denoted by $\YoungmAA$ Young diagram. In this case we choose $\ttt_{AB}=-\ttt_{BA}$ and, hence, for the gauge field we find
\begin{align}
    \omega&= \omega^{a+}\ttt_{a+} +\tfrac12 \omega^{a,b}\ttt_{ab}+\omega^{+-}\ttt_{+-}+\omega^{a-}\ttt_{a-}\label{eq317}
\end{align}
and similarly for the gauge parameter $\xi$. The action of $\PPP_a$ is read off from \eqref{Tt}:
\begin{align*}
    [\PPP_a, \ttt_{c+}]&=0\,, & [\PPP_a, \ttt_{c-}]&=-\ttt_{ac}-\eta_{ac}\ttt_{+-}\,, & [\PPP_a, \ttt_{+-}]&=-\ttt_{a+}\,, & [\PPP_a, \ttt_{cd}]&=\eta_{ac}\ttt_{d+}-\eta_{ad}\ttt_{c+} \,.
\end{align*}
The linearized conformal gravity equations and gauge symmetries are
\besubeqs
\begin{align}
    \ttt_{+-}&: &d\omega^{+-}-h_m \wedge \omega^{m-}&=0\,, &\delta \omega^{+-}&=d\xi^{+-}-h_m \xi^{m-}\,,\\ 
    \ttt_{a+}&: &d\omega^{a+}+h_m \wedge \omega^{m,a}-h^a\wedge \omega^{+-}&=0\,, &\delta \omega^{a+}&=d\xi^{a+}+h_m \xi^{m,a}-h^{a}\xi^{+-}\,,\\
    \ttt_{ab}&: &d\omega^{a,b}-h^a \wedge \omega^{b-}+h^b \wedge \omega^{a-}&=0\,, &\delta \omega^{a,b}&=d\xi^{a,b}-h^a \xi^{b-}+h^b \xi^{a-}\,,\\ 
    \ttt_{a-}&: &d\omega^{a-}&= 0\,, &\delta \omega^{a-}&=d\xi^{a-} \,.
\end{align}
\esubeqs
The analysis is, of course, the linearized version of the one in section \ref{sec:review}.\footnote{In flat space and Cartesian coordinates, we identify world indices $\mm,\nn,...$ and fiber ones $a,b,...$. In what follows we implicitly convert the world indices into the fiber ones with the help of $h_{\mm }^a=\delta^a_\mm$ and write them after the separator $|$, e.g. $\omega^{ab|m}\equiv \omega^{ab}_\mm\, h^{\mm d}$. } (1) $\omega^{+-}$ can be gauged away with the help of $\xi^{m-}$, the leftover gauge symmetry being with $\xi^{m-}=\pl^m \xi^{+-}$; (2) the first equation implies $dx^\mm \wedge dx^\nn\, h_{c\nn}\,\omega^{c-}_\nn=0$, i.e. the anti-symmetric component of $\omega^{c-|d}\equiv \omega^{c-}_\mm h^{\mm d}$ must vanish, so that $\omega^{c-|m}=S^{cm}$ for some symmetric $S^{ab}$; (2) as in the gravity case, we can use the local Lorentz symmetry $\xi^{a,b}$ to eliminate the antisymmetric part of the dynamical dreibein $\omega^{a+|m}$, so that $\omega^{a+|m}=\phi^{am}$ for some symmetric $\phi^{ab}$; (3) the algebraic gauge symmetry $\delta \phi^{ab}=-\eta^{ab}\xi^{+-}$ with $\xi^{+-}$ allows us to gauge away the trace, so that $\phi^{ab}$ is traceless; (4) the second equation allows us to solve for $\omega^{ab|m}=-\pl^a\phi^{bm}+\pl^b\phi^{am}$; (5) the third equation expresses $S^{ab}$ as the linearized Schouten tensor 
\begin{align*}
    S_{ab}=\frac12 (-\square \phi_{ab} +\pl_a \pl^m\phi_{mb}+\pl_b \pl^m\phi_{ma}-\tfrac12 \eta_{ab} \pl^m\pl^n \phi_{mn})\,.
\end{align*}
(6) so far all the equations merely expressed fields as derivatives of some other fields. The only dynamical equation is the last one that sets to zero the Cotton tensor
\begin{align}
    C_{ab}&=\epsilon\fdu{a}{mn}\pl_m S_{bn}+\epsilon\fdu{b}{mn}\pl_m S_{an}= 0\,, & \delta \phi_{ab}&=\pl_a\xi_b +\pl_b \xi_a -\frac{2}{3}\eta_{ab} \pl^m\xi_m\,.
\end{align}

\paragraph{Depth-two Conformal Gravity.} The simplest new case is to take a vector representation of $so(3,2)$, denoted by $\YoungA$, i.e. we have $\ttt_{A}$ as a base. The gauge field decomposes as
\begin{align}
    \omega&= \omega^{a}\ttt_a +\omega^+ \ttt_++\omega^{-} \ttt_-\,.
\end{align}
The action of the translation generators $\PPP_a$ reads
\begin{align}
    [\PPP_a,\ttt_+]&=0\,, & [\PPP_a,\ttt_-]&=\ttt_a\,,& [\PPP_a,\ttt_c]&=-\eta_{ac}\ttt_+\,.
\end{align}
The linearized system of equations and gauge symmetries is then
\besubeqs
\begin{align}
    \ttt_+&: & d \omega^+ -h_m\wedge \omega^m &=0\,, & \delta \omega^+&= d\xi^+-h_m \xi^m\,,\\
    \ttt_c&: &d\omega^c +h^c\wedge \omega^-&=0\,, & \delta \omega^c&= d\xi^c+h^c \xi^-\,,\\
    \ttt_-&: & d \omega^-&= 0\,, & \delta \omega^-&= d\xi^-\,.
\end{align}
\esubeqs
Similarly to the partially-massless case \cite{Skvortsov:2006at}, one proceeds as follows. With the help of $\xi^c$ we can gauge away $\omega^+$. The leftover gauge transformations obey $\pl_m \xi^+ - \xi_m=0$. The first equation implies that the anti-symmetric component of $\omega^c=\omega^c_\mm \, dx^\mm$ vanishes, i.e. we can identify $\omega^{c|m}=\phi^{\mm c}$ for some symmetric $\phi_{ab}$. The $\xi^-$ gauge symmetry $\delta \phi_{ab}=\pl_a \xi_b+\eta_{ab}\xi^-$ can be used to make $\phi_{ab}$ traceless. As a result we are left with ($\xi\equiv \xi^+$)
\begin{align}\label{spintwog}
    \delta \phi_{ab}=\pl_a \pl_b\xi -\frac{1}{3}\eta_{ab} \square \xi\,.
\end{align}
This is the depth-two Fradkin-Tseytlin spin-two conformal field.\footnote{This was to some extent studied in \cite{Pope:1989vj}, but there this mode was identified as a 'non-gauge spin-two field'. We show that it is a gauge field, in fact. } The second equation can be projected onto the two irreducible components, one of them is used to solve for $\omega^-_m$ and another one leads to the Cotton tensor
\begin{align}
    C_{ab}=\epsilon^{mn}{}_a\partial_{m}\phi_{bn}+\epsilon^{mn}{}_b\partial_{m}\phi_{an}= 0\,.
\end{align}
that obeys the Noether identity $\pl^a \pl^b C_{ab}\equiv 0$ as a consequence of \eqref{spintwog}.

\paragraph{Depth-three spin-three field.} The simplest higher spin example is to take an irreducible rank-two symmetric representation of $so(3,2)$ denoted by $\YoungpB$, i.e. $\ttt_{AB}=\ttt_{BA}$, $\ttt_{AB}\eta^{AB}=0$. The decomposition of $\omega$ can be chosen as
\begin{align}
    \omega&= \omega^{a+}\ttt_{a+} +\tfrac12 \omega^{ab}\ttt_{ab}+\frac12\omega^{++}\ttt_{++}+\frac12\omega^{--}\ttt_{--}+\omega^{a-}\ttt_{a-}\,,
\end{align} 
where we took into account the tracelessness, $\ttt\fud{m}{m}+2\ttt_{+-}=0$. Note that $\omega^{ab}$ is not traceless, but we do not have to introduce $\omega^{+-}$. The action of $\PPP_a$ is
\begin{align*}
    [\PPP_a, \ttt_{c+}]&=-\eta_{ac}\ttt_{++}\,, & [\PPP_a, \ttt_{c-}]&=\ttt_{ac}+\tfrac12t\fdu{m}{m}\,, & [\PPP_a, \ttt_{--}]&=2\ttt_{a-}\,, \\ [\PPP_a, \ttt_{++}]&=0\,, &
    [\PPP_a, \ttt_{cd}]&=-\eta_{ac}\ttt_{d+}-\eta_{ad}\ttt_{c+}\,. 
\end{align*}
The linearized equations and gauge symmetries are
\begin{align*}
    \ttt_{++}&: &d\omega^{++}-2h_m\wedge\omega^{m+}&=0\,, &\delta\omega^{++}&= d\xi^{++}-2h_m\xi^{m+}\,,\\
    \ttt_{a+}&: &d\omega^{a+}-h_m \wedge \omega^{ma}&=0\,, &\delta \omega^{a+}&=d\xi^{a+}-h_m \xi^{ma}\,,\\
    \ttt_{ab}&: &d\omega^{ab}+h^{(a} \wedge \omega^{b)-}+h_m\wedge\omega^{m-}\eta^{ab}&=0\,, &\delta \omega^{ab}&=d\xi^{ab}+h^{(a} \xi^{b)-}+h_m\xi^{m-}\eta^{ab}\,,\\ 
    \ttt_{a-}&: &d\omega^{a-}+h^a\wedge\omega^{--}&= 0\,, &\delta \omega^{a-}&=d\xi^{a-}+h^a\xi^{--}\,, \\
     \ttt_{--}&: &d\omega^{--}&=0\,, & \delta\omega^{--}&=d\xi^{--}\,.
\end{align*}
Here, we start by setting $\delta \omega^{++}=0$ and the leftover gauge symmetry requires $\xi^{m+}=\tfrac12 \pl^m\xi^{++}$. The $\ttt_{++}$-equation implies that $\omega^{m+|n}$ is symmetric. Proceeding to $\ttt_{a+}$ we observe that symmetric $\omega^{m+|n}$ can be eliminated with the help of $\xi^{ab}$, i.e. $\omega^{a+}=0$ now. The leftover gauge symmetry requires $\xi^{ab}=\tfrac12 \pl^a\pl^b \xi^{++}$. The $\omega^{a+}$ equation of motion implies that $\omega^{ab|m}=\phi^{abm}$ for some symmetric $\phi^{abc}$. At the next step, $\ttt_{ab}$, the $\xi^{a-}$ gauge symmetry allows us to make $\phi^{abc}$ traceless. Therefore, we see that $\phi^{abc}$ transforms as ($\xi^{++}=2\xi$)
\begin{align}\label{spinthreeg}
    \delta \phi^{abc}&= \pl^a\pl^b\pl^c \xi -\frac15(\eta^{ab}\pl^c \square \xi+\eta^{ac}\pl^b \square \xi+\eta^{ac}\pl^b \square \xi)\,.
\end{align}
i.e. it is a depth-three Fradkin-Tseytlin field. The $\omega^{ab}$ equation of motion has several $so(2,1)$-irreducible components. Some of them allows to solve for $\omega^{a-}$, while there is one that leads to the dynamical equation 
\begin{align}
    C^{abc}&=\epsilon^{amn}\pl_m \phi\fdu{n}{bc}+\epsilon^{bmn}\pl_m \phi\fdu{n}{ac}+\epsilon^{cmn}\pl_m \phi\fdu{n}{ab}=0\,,
\end{align}
where $C^{abc}$ is the Cotton tensor. The equation for $\omega^{a-}$ allows us to solve for $\omega^{--}$ and imposes no additional constraints on $\phi$. One can also find the Noether identity $\pl^{a} \pl^b \pl^c C_{abc} \equiv0$ manifesting the gauge symmetry \eqref{spinthreeg}. 

\paragraph{General case.} In the general case, given a higher spin algebra $\hs$ with an embedding of $so(3,2)$ we need to decompose $\hs$ into $so(3,2)$-modules. The spectrum of Fradkin-Tseytlin fields is determined by the free field limit.  As is anticipated in the introduction the complete dictionary between $so(3,2)$-modules and conformal fields reads\footnote{In practice, one needs to replace the words depth-$t$ partially-massless fields with depth-$t$ Fradkin-Tseytlin fields in the dictionary of \cite{Skvortsov:2006at}. Rigorous analysis can be done by dimensional reduction of \cite{Skvortsov:2009nv} or by redoing  \cite{Vasiliev:2009ck, Bekaert:2013zya} in $3d$. }
\begin{align}
    \delta \phi_{a_1...a_s} &= \pl_{a_1}...\pl_{a_t}\xi_{a_{t+1}...a_s}+... && \Longleftrightarrow && \parbox{90pt}{\begin{picture}(90,20)\put(0,0){\RectBRow{9}{7}{$s-1$}{$s-t$}}\put(0,0){\YoungAA}\put(60,0){\YoungA}\put(80,10){\YoungA}\end{picture}}
\end{align}
Here, the gauge field is $\omega^{A(s-1),B(s-t)}$ and gauge parameter is $\zeta^{A(s-1),B(s-t)}$. It is quite easy to see that all gauge parameters except for $\zeta^{a(s-t)+(t-1),+(s-t)}$ can be used to gauge away certain components of $\omega^{A(s-1),B(s-t)}$. Here, ${+(k)}$ denotes $+...+$  ($k$ times). All components of $\omega$ except for
\begin{align}
    \phi^{a_1...a_s}&=\omega^{a_1...a_{s-1},+(s-t)}_\mm h^{\mm a_s}+\text{symmetrization}-\text{traces}
\end{align} play the role of generalized auxiliary fields (can be gauged away or solved for). The dynamical equations of motion can be written in terms of the Cotton tensor that is an operator of order $2s-2t+1$ that is symmetric and traceless
\begin{align}
    C_{s}&= \epsilon^{...} \pl^{2s-2t+1}\phi_s\,.
\end{align}
Also, $C_{a_1...a_s}$ contains $\epsilon^{abc}$. It obeys the Noether identity $\pl^{b_1}...\pl^{b_t} C_{b_1...b_t a_{t+1}...a_s}\equiv0$. The higher derivative equations are obtained by solving for a number of auxiliary fields in terms of $\phi_{a_1...a_s}$, one by one. They reside in one component of the $\ttt_{a_1...a_{s-1},-(s-t)}$-equation.

Note that the $3d$ spinorial language is very helpful in dealing with the conformal higher spin fields. Every rank-$s$ tensor $\phi_{a_1...a_s}$ corresponds to a rank-$2s$ $sl_2$-tensor $\phi_{\ga_1...\ga_{2s}}$. Here, $\ga,\gb,...=1,2$ are the indices of $sl_2$. The dictionary between the $sl_2$-base and the $so(2,1)$-base is via Pauli matrices, $\sigma_m^{\ga\gb}$. The gauge transformations for the depth-$t$ Fradkin-Tseytlin fields are simply
\begin{align}
    \delta\phi_{\ga(2s)}&= \overbrace{\pl_{\ga\ga}...\pl_{\ga\ga}}^t \xi_{\ga(2s-2t)}\,,
\end{align}
where the symmetrization is implied. The advantage of the spinorial language is that the complicated projector onto the traceless part is not needed. 

\subsection{Comments on Interacting Fradkin-Tseytlin Fields}
\label{sec:interactions}
Let us elaborate more on the formulation of the conformal higher spin gravities in terms of Fradkin-Tseytlin fields. The linearized gauge symmetries are
\begin{align}
    \delta \phi_{a_1...a_s} &= \pl_{a_1}...\pl_{a_t}\xi_{a_{t+1}...a_s}+\text{permutations}-\text{traces}\,.
\end{align}
Such fields naturally couple to the (partially)-conserved tensors \cite{Dolan:2001ih}
\begin{align}
    \pl^{b_1}...\pl^{b_t} J_{b_1...b_t a_{t+1}...a_s}&=0
\end{align}
that one finds in higher derivative free conformal field theories of type $\square^k \Phi=0$ or $\square^{k-1}\slashed \pl \Psi=0$.\footnote{In such theories one finds partially-conserved tensors of odd depths. By taking free CFT's that contain free fields with different $k$'s one can find tensors with even $t$. } The same time, such fields are boundary values of partially-massless fields \cite{Dolan:2001ih, Bekaert:2013zya}. The currents have conformal weight $s-t+2$, hence the weight of $\phi_{a(s)}$ is $t+1-s$ and the weight of $\xi_{a(t)}$ is $1-s$. Therefore, when the action is expressed in terms of Fradkin-Tseytlin fields $\phi_{s,t}$ we expect to find 
\begin{align}
    S=\int d^3x\, \epsilon^{...} \left( \sum_{s,t} \phi_{s,t}\,  \pl^{2s-2t+1}\phi_{s,t} +\sum_{s_i,t_i}\pl^{N_{s_i,t_i}}\phi_{s_1,t_1}\phi_{s_2,t_2}\phi_{s_3,t_3}+...\right)\,,
\end{align}
where the number of derivatives $N_{s_i,t_i}$ in the vertex is fixed to be $3+\sum_i(s_i-t_i+1)$. Similarly, the gauge transformations receive higher order corrections. The important feature of conformal higher spin theories is that the number of derivatives in a vertex is fixed for any given $s_i,t_i$. Therefore, conformal higher spin theories are always perturbatively local. The equations of motion should give the higher spin Cotton tensor (an appropriate multiplet of Cotton tensors, as prescribed by a given higher spin algebra).

It would be important to understand the higher spin geometry that underlies conformal higher spin theories and Cotton tensors. Another motivation to resort to the metric-like fields is to discuss matter couplings, see e.g. \cite{Nilsson:2013tva,Nilsson:2015pua,Linander:2016brv}. It is unclear at present how to couple matter fields directly to the Chern-Simons formulation even in the conformal gravity case, but is trivial, of course, in the metric-like formulation. 

There is no conceptual problem in getting the nonlinear theory of $3d$ Fradkin-Tseytlin fields from the Chern-Simons formulation, which is similar to the massless case \cite{Fredenhagen:2014oua}. In the weak field expansion for every given order $n$ one finds the system
\begin{align}
d\omega_n +\omega_0\wedge \omega_n+\omega_n\wedge \omega&=-\sum_{\substack{i+j=n\\ i,j>0}} \omega_i\wedge \omega_j\,,
\end{align}
where the lower order fields $\omega_i$ are already expressed in terms of Fradkin-Tseytlin fields and derivatives thereof. At order-$n$ one finds non-vanishing 'torsion' on the right-hand side. Upon expressing various components $\omega_n$ in terms of $\phi$'s at order-$n$ the torsion can be absorbed into the solution for $\omega_n$. The only component that cannot be absorbed, as is clear from the linearized analysis, is the Cotton tensor $C_n$ (to be precise a collection of tensors that depends on the spectrum of fields). 

This way one can get, as a matter of principle, the order-by-order expression for the Cotton tensor(s). We expect that this problem is much simpler than the one for massless fields in anti-de Sitter space since conformal fields can be studied in their simplest background, i.e. the flat space, and derivatives in the interaction vertices are constrained by the conformal weight. Indeed, in the simplest nontrivial case of $\hs=so(4,2)$, i.e. $sl(V)$ for $V$ being the spinorial representation of $so(3,2)$, the complete non-linear theory was derived in \cite{Pope:1989vj}.

Another way to approach the problem is the bottom-up Noether procedure, i.e. to try to construct order-by-order a consistent non-linear theory for a given spectrum of fields. It is tempting to conjecture that we have identified all such basic theories and they are in one-to-one with the list of higher spin algebras given above. The same time, given the simplicity of the conformal fields as compared to their massless cousins, the Noether procedure might be completed. 

We would like to recall that the conformal gravity admits three equivalent formulations: (a) as the Chern-Simons action for the conformal algebra $so(3,2)$; (b) as a hybrid Chern-Simons action for the Lorentz algebra $so(2,1)$ where the spin-connection is assumed to be solved for from the torsion constraint; (c) the formulation where the dynamical variable is the conformal metric $g_{\mm\nn}$. While Fradkin-Tseytlin fields lead clearly to (c), it would be interesting to see if there exists an intermediate hybrid formulation of type (b) in the higher spin case as well. 

Any of the conformal higher spin theories that we discussed contain the conformal graviton by construction. There are exact solutions where all $s>2$ modes are switched off --- conformally flat backgrounds. Each theory can be linearized over such a background and we observe a number of non-interacting Fradkin-Tseytlin fields. This shows that the Fradkin-Tseytlin fields can propagate on conformally flat backgrounds. An explicit component form of the actions for Fradkin-Tseytlin fields on conformally flat backgrounds was obtained in \cite{Kuzenko:2019ill}.

\section{Conformal higher spin fields as background fields}
\label{sec:background}
A standard way to arrive at off-shell Fradkin-Tseytlin fields is to start with a matter fields and to consider a theory of its associated background fields. More specifically, taking for simplicity the scalar field one can write its action on a general higher spin background as~\cite{Segal:2002gd}
\begin{align}
    S=\int d^dx\, \phi^*(x) H(x,\pl)\phi(x)\,, \qquad H=H_0(x)+H^a(x)\d_a+H^{ab}(x)\d_a\d_b+\ldots\,.
\end{align}
Natural gauge symmetries of this action are given by
\begin{align}\label{gaugeback}
\delta H &= HU+U^\dagger H\,, && \delta \phi=-U\phi\,, && U=U(x,\pl)\,.
\end{align}
This construction has a direct extension to more general systems including spinning conformal fields in terms of the respective quantum constrained systems~\cite{Grigoriev:2006tt,Bekaert:2013zya}. In $d>3$ this nonlinear symmetry is a starting point in constructing both off-shell and on-shell conformal theories in even dimensions as well as associated massless higher spin theory in $AdS_{d+1}$ \cite{Bekaert:2017bpy,Grigoriev:2018wrx}. In particular, the higher spin algebra arises as a stability algebra of the vacuum $H_{vac}$. This algebra is also known as that of global reducibility parameters. For instance, in the case of scalar field taking $H_{vac}=(\eta^{ab}\d_a\d_b)^k$ one arrives at the usual (partially-massless) higher spin algebras discussed in section \ref{sec:hsa}.

In $3d$ the above strategy leads to a nonlinear theory of off-shell Fradkin-Tseytlin fields. In order to derive an on-shell theory one can try first to analyze the possible linear equations. Using the known classification of conformally invariant equations~\cite{Shaynkman:2004vu} one concludes that there are only two options: either to leave higher spin Cotton tensor associated to $H_{a_1\ldots a_s}(x)$ unconstrained or set it to zero. The former option leaves us with the off-shell theory while the latter leads to Chern-Simons theory discussed in the previous section. Indeed, at the linearized level this immediately follows from the analysis of section~\ref{sec:freefields}. If in addition we assume that the nonlinear theory should be a consistent deformation of the linearized one it is enough to work in the Chern-Simons formulation, where the only fields are a one-form $\omega$ that takes values in a higher spin algebra (seen as a linear space). This theory is of AKSZ type \cite{Alexandrov:1995kv} and any its deformation should be also of AKSZ type \cite{Barnich:2009jy} and hence determined by a Lie algebra structure on the space where one-form fields take values. Consistency with the algebra of global reducibility parameters requires that the algebras should coincide. 

Despite the above logic gives the first-principle derivation of the conformal higher spin Chern-Simons theory in the case of usual infinite-dimensional higher spin algebras, it does not cover the case of finite-dimensional ones.  It turns out that this can be done in a similar way but with the conformal scalar replaced by topological matter that does not carry local degrees of freedom.

Conformally-invariant matter equations of motion on a conformally-flat background (e.g. Minkowski space) can always be represented as a covariant constancy condition where the connection is associated to a flat Cartan connection describing the space-time geometry.  Let $V$ be the space of locally defined solutions, which is a module over $so(3,2)$  thanks to conformal invariance. The matter equations of motion take the form
\begin{equation}\label{covconst}
(d+\rho(\omega))\phi=0\,,
\end{equation}
where $\rho:so(3,2)\to \mathrm{End}(V)$ denotes a representation map and $\phi$ is a $V$-valued function (in general it is a section of a vector bundle with typical fiber $V$).

A general background field for such a matter field is simply a generic linear flat connection $A$ valued in $\mathrm{End}(V)$. One can in principle consider non-flat connections but this would effectively reduce $V$ to the kernel of the curvature and we exclude such more general and involved situations. The natural gauge transformations for this system are
\begin{align}
\delta A&=d\epsilon+[{A},{\epsilon}]\,, && \delta \phi=-\epsilon \phi\,.
\end{align}
It is clear that algebra of vacuum symmetries coincides with $\mathrm{End}(V)$.

To conclude, the natural background fields for topological matter are flat  $\mathrm{End}(V)$-connections. This gives an independent argument justifying that the Chern-Simons theory for $\mathrm{End}(V)$ can be derived from the topological matter system pretty much the same way as the ordinary conformal higher spin gravity is derived from the standard $3d$ singleton. The important difference, however, is that in the case of topological matter neither we encounter off-shell conformal fields (for which Cotton tensor may be non-vanishing) nor we can write an action for $\phi$ that leads to \eqref{covconst}.

\section*{Acknowledgments}
\label{sec:Aknowledgements}
We are grateful to Kostya Alkalaev, Thomas Basile, Xavier Bekaert and Sergei Kuzenko for very useful discussions. The work of M.G. was supported by RFBR grant No. 18-02-01024. The work of I.L. was supported by the
grant J4129-N27 of the Austrian Science Fund (FWF) and by the ST/P000762/1 grant of Science and Technology Facilities Council (STFC). M.G. and E.S. are grateful to the organizers of APCTP-KHU Workshop on Higher Spin Gravity where this work was finished and presented.

\footnotesize
\providecommand{\href}[2]{#2}\begingroup\raggedright\endgroup

\end{document}